\documentclass[aps,prb,12pt,notitlepage,tightenlines,showpacs,showkeys]{revtex4-1}
\usepackage{graphicx}
\usepackage{amsmath,amssymb}

\newcommand{\PR}{{\em Phys. Rev. }}
\newcommand{\PRL}{{\em Phys. Rev. Lett. }}
\newcommand{\PL}{{\em Phys. Lett. }}
\newcommand{\NP}{{\em Nucl. Phys. }}
\newcommand{\etal}{{\em et al}}
\newcommand{\AP}{{A$^\prime$}}

\def\AB{\mbox{A$^\prime$-boson}}
\def\V3{\mbox{VEPP--3}}

\begin{document}
\title{Searching for a new force at \V3}
\author{B.~Wojtsekhowski}
\email{E-mail: bogdanw@jlab.org}
\affiliation{Thomas Jefferson National Accelerator Facility, 
Newport News, VA 23606, USA }
\author{D.~Nikolenko}
\author{I.~Rachek}
\affiliation{Budker Institute of Nuclear Physics of Siberian 
Branch Russian Academy of Sciences, 
630090 Novosibirsk, Russia}
\begin{abstract}
We propose an experiment to search for a new gauge boson, \AP, in $e^+e^-$ annihilation
by means of a positron beam incident on a gas hydrogen target internal
to the \V3 storage ring.  
The search method is based on a missing mass spectra in the reaction
$e^+e^-\rightarrow \gamma$\AP.
It allows observation of the \AP~signal independently of its decay modes and life time.
The projected result of this experiment corresponds to an upper limit on the square
of coupling constant $|f_{e\text{A}{^\prime}}|^2=1\cdot 10^{-8}$ with a signal-to-noise ratio of five
to one at an \AP~mass of 15 MeV.
\end{abstract}


\maketitle

\section{Introduction}
The search for an experimental signature of physics beyond the Standard Model
is a major effort of modern particle physics, see e.g.~\cite{pdg4}.
Most of the search activity is focused on possible heavy particles with 
a mass scale of 1~TeV and above.  
At the same time, as was suggested by P.~Fayet~\cite{fa80,fa90}, there could be  
extra $\text{U(1)}$ symmetry, which requires a new gauge boson, U,
also called the \AB.  
The boson could be light and weakly interacting with known particles.
Most constraints for the light \AB~parameters were obtained from
electron and muon anomalous magnetic moments $(g-2)$ and 
particle decay modes \cite{fa06,fa07,fa08,po09}.

Renewed interest in a search for the new gauge boson has been seen recently as 
such a boson may provide an explanation for various astrophysics phenomena, 
accumulated during the last decade,  which are related to dark matter %
\cite{fa06,fa07,arkani,est09}.  
The possible connection between the \AB~and dark matter in view of the
observed slow positron abundance has been investigated for several years 
and is often referred to as MeV dark matter (MDM)\cite{bo04,fa04,bofa,bo06}.  
The theory of dark matter proposed by N.~Arkani-Hamed and
collaborators~\cite{arkani}, which provided an interpretation of a number
of key astrophysical observations, 
sparked additional interest in a \AB~search in the mass range below 1~GeV.  

Several methods were used in the search for the \AB~signal,
considering ``invisible'' decay modes of the \AB.  Therefore, they result 
in an upper limit for ${f_{_{{qA}{^\prime}}} \times B_{[A{^\prime}\rightarrow    invisible]}}$.  
The first method uses precise experimental data on exotic
decay modes of elementary particles, e.g. 
\mbox{$\pi^\circ \rightarrow invisible + \gamma$},
for the calculation of the upper limit of the \AB~coupling constant 
to the specific flavor.  
These upper limits for decay of the $J/\Psi$ and $\Upsilon$ to a photon
plus invisible particles were obtained experimentally
by means of the ``missing particle'' approach, where 
a missing particle in the event type $e^+e^- \rightarrow \gamma X$
leads to a yield of events with a large energy photon detected at 
a large angle with respect to the direction of positron 
and electron beams.  
From the yield of such events the coupling constant could be determined
for a wide range of mass of the hypothetical \AB.  
A recent experiment~\cite{BABAR} using statistics of 
$1.2\cdot 10^8 \,\,\, \Upsilon(3S)$ events provided the best
data for $\Upsilon(3S)$ decay to $A^0 + \gamma$ and a 
limit on the coupling of the \AB~to the $b$-quark.
In the mass region below 100~MeV, 
the limit for 
$B(\Upsilon(3S) \rightarrow \gamma A^0) \times B(A^0 \rightarrow 
\text{invisible})$ is $3 \cdot 10^{-6}$, 
here the $A^0$ notation used as in paper~\cite{BABAR},
from which the limit ${{f_{_{bA^0}}} < 4 \cdot 10^{-7} m_{_{A^0}\,[\mathrm {MeV}]}}$ 
was obtained~\cite{fa08}.  
An additional hypothesis of coupling constants universality is required 
to get a bound on ${f_{_{{eA}{^0}}}}$, so direct measurement 
of the coupling to an electron is of large interest.  
Currently, the upper limit on the vector coupling 
obtained from the discrepancy between the calculated electron anomalous magnetic moment
and the measured one is ${f_{_{{e{\text A}^\prime}}} < 1.0 \cdot 10^{-4} \, 
\cdot m_{_{{\text A}^\prime}}\,[\text{MeV}]}$~\cite{fa07,po09}.

A direct measurement of $f_{_{eA'}}$ and $\mathbf {m_{_{A'}}}$ could be made by detecting 
the decay of the \AB~to an electron-positron pair and reconstructing 
the $e^+e^-$ invariant mass.  
It requires a significant branching of \AB~decay to the $e^+e^-$ pair.
A complication of this method is the high level of 
electromagnetic background in the mass spectrum of $e^+e^-$, 
so such a measurement requires large statistics.
Recently the data sets accumulated in collider experiments  
have been used for such an analysis \cite{bo06,babar,kloe}.

Electron fixed-target experiments, where a new boson can be produced from radiation off
an electron beam incident on an external target, are now widely
discussed \cite{ap2,ap3,ap4,apex1}.  
The first significant experimental results on upper limits for a new boson coupling to an electron
in the sub-GeV mass range have been reported \cite{mainz,apres}.  
The APEX experiment in JLab Hall A \cite{apex2} will probe couplings 
$\alpha'/\alpha > 10^{-7}$ and masses $\sim 50 - 550$~MeV.  
The result of the test run, with only 1/200 of the data of the full APEX experiment,
has already demonstrated the feasibility of such an approach~\cite{apres}.  
Other electron fixed-target experiments are planned: \  at Jefferson Lab,
including the Heavy Photon Search (HPS) \cite{HPS} and DarkLight \cite{darklight};
at MAMI \cite{mainz}. 

A sensitive \AB~search could be performed with a low energy 
$e^+e^-$ collider \cite{est09}, where several search techniques could be used:
\begin{itemize}
\item The invisible particle method
\item Invariant mass of the final $e^+e^-$ pair
\item Missing mass with single-arm photon detection. 
\end{itemize}

To search for the \AB~with a mass of 10-20~MeV, 
the center-of-mass energy of $e^+e^-$, $E_{cm}$, should be low.
The production cross section is proportional to $1/E_{cm}^2$, so for 
low $E_{cm}$, even a modest luminosity would be sufficient for 
a precision measurement.  

However, no colliding ($e^-e^+$)--beam facility in this energy region 
exists or is planned to be constructed.
Still, a similar operation can be achieved if an available positron beam of
a few hundred MeV energy is incident on a fixed target \cite{wo06, ref:concept}. 
The \V3 electron/positron storage ring at the Budker Institute at Novosibirsk \cite{vepp}, 
with its internal target facility and high-intensity positron beam injection complex, 
are uniquely suited for such measurements.
\vskip+\baselineskip

{\bf
We propose to perform a search for the \AB~in a mass range 
$\mathbf{m_{_{{\text A}^\prime}}=5-20}$~MeV using 
a 500~MeV positron beam incident on an internal hydrogen target, providing a
luminosity of $\mathbf{10^{32}}$~cm$\mathbf{^{-2}}$s$\mathbf{^{-1}}$, by detecting 
$\mathbf{{\gamma}}$-quanta 
from the process $\mathbf{e^+e^-\rightarrow \gamma {\text A}^\prime}$ 
in an energy range  $\mathbf{E_\gamma=50-400}$~MeV and angular range 
$\mathbf{\theta_\gamma^{CM}=90^\circ\pm 30^\circ}$ 
($\mathbf{\theta_\gamma^{Lab}=1.5^\circ - 4.5^\circ}$).
}

\section{The concept}
In the proposed experiment we would like to explore the technique of the missing mass
measurement approach with single-arm photon detection. 
The concept of the method is partly described in \cite{ref:concept}.
A positron beam in a storage ring with an energy $E_+$ of a few hundred MeV and an internal
hydrogen gas target make up an ``$e^+e^-$ collider''. In such a collider it is possible
to search for a light \AB~with a mass of up to $m_{{\text A}^\prime}[MeV]\sim \sqrt{E_+[MeV]}$.
Unlike all other experiments with a fixed target, which are based on the detecting of $e^-e^+$ pairs
from \AB~decay, 
\textbf{\textit{in the proposed experiment  no special assumptions about decay modes 
of the \AB~are required.}} 
In this proposal we consider a low luminosity ($\sim 10^{32}$~cm$^{-2}$s$^{-1}$) 
measurement and a combination of on-line and off-line veto on the bremsstrahlung 
and multi-photon background processes.

\begin{figure*}[htb]
  \includegraphics[width=0.7\linewidth]{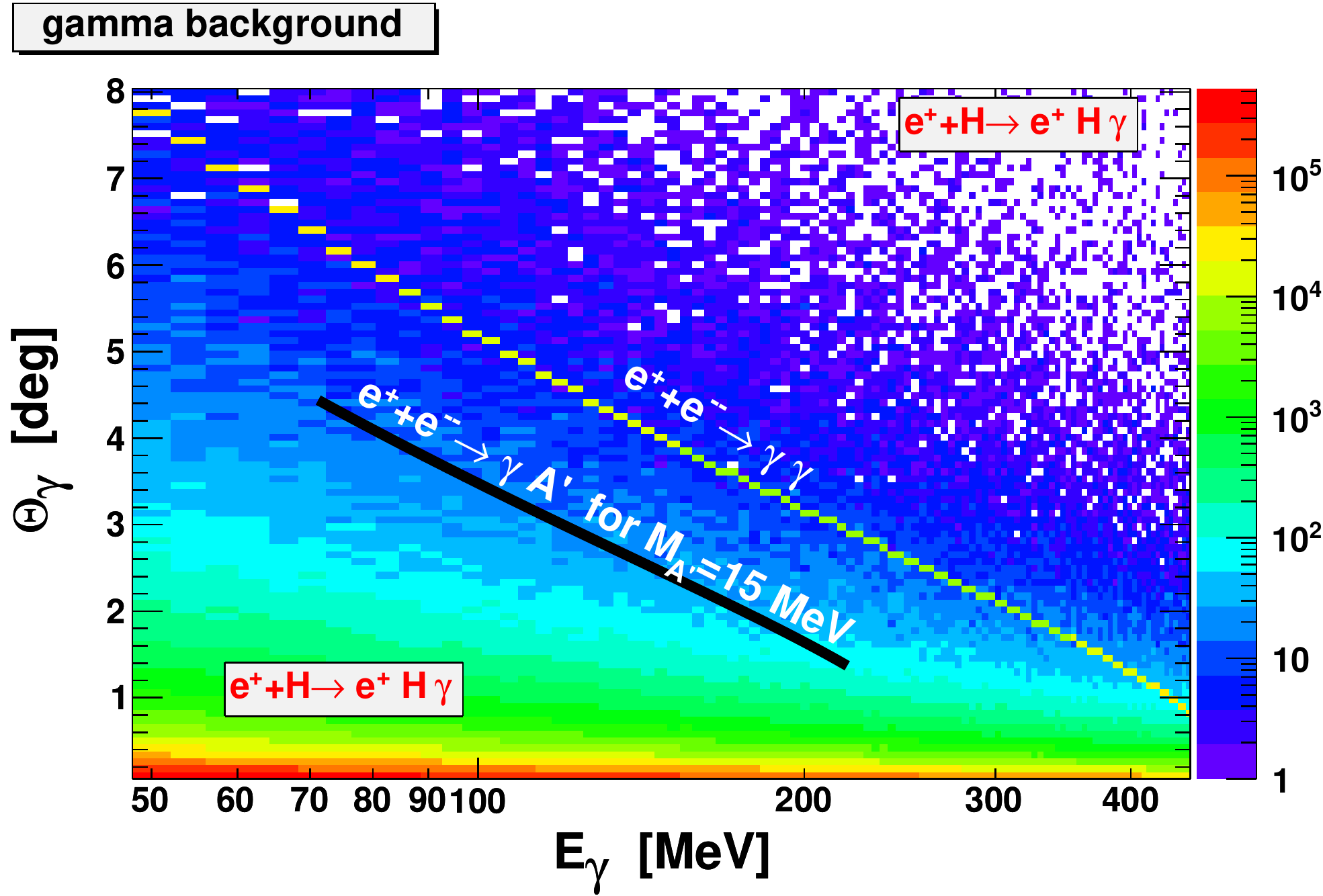}
\caption[]{\label{fig:2d}
Two-dimensional distribution of the photon events in the scattering angle and the 
photon energy for a 500 MeV positron beam incident on a hydrogen gas target. 
The black band shows the location of \AB~events of 15 MeV mass. 
}
\end{figure*}

In the process $e^+e^-\rightarrow {\text A}^\prime\gamma$ a measurement of the photon energy and its 
angle allows a reconstruction of the missing mass spectrum and a search for a peak
corresponding to the \AB. In such a spectrum the dominant signal corresponds to the 
annihilation reaction $e^+e^-\rightarrow \gamma\gamma$. The signal for the \AB~will be
shifted to the area of the continuum (see the illustration in Fig.~\ref{fig:2d}).
The continuum part of the event distribution is dominated by photons emitted 
in a process of positron scattering from an electron or a proton in the target 
(bremsstrahlung) and by photons from the three-photon annihilation process. 
Contributions of other reactions, e.g. 
$\gamma^\star p \rightarrow p \pi^0 \rightarrow p \gamma\gamma$,
are at least three orders of magnitude smaller than that of positron bremsstrahlung.

A key property of the proposed experimental setup is the ability to suppress the QED background 
significantly, both on-line and off-line, thus greatly improving the sensitivity of the search. 

\section{The kinematics and cross section}
\begin{figure}[htbp]
        \includegraphics[width=0.45\linewidth]{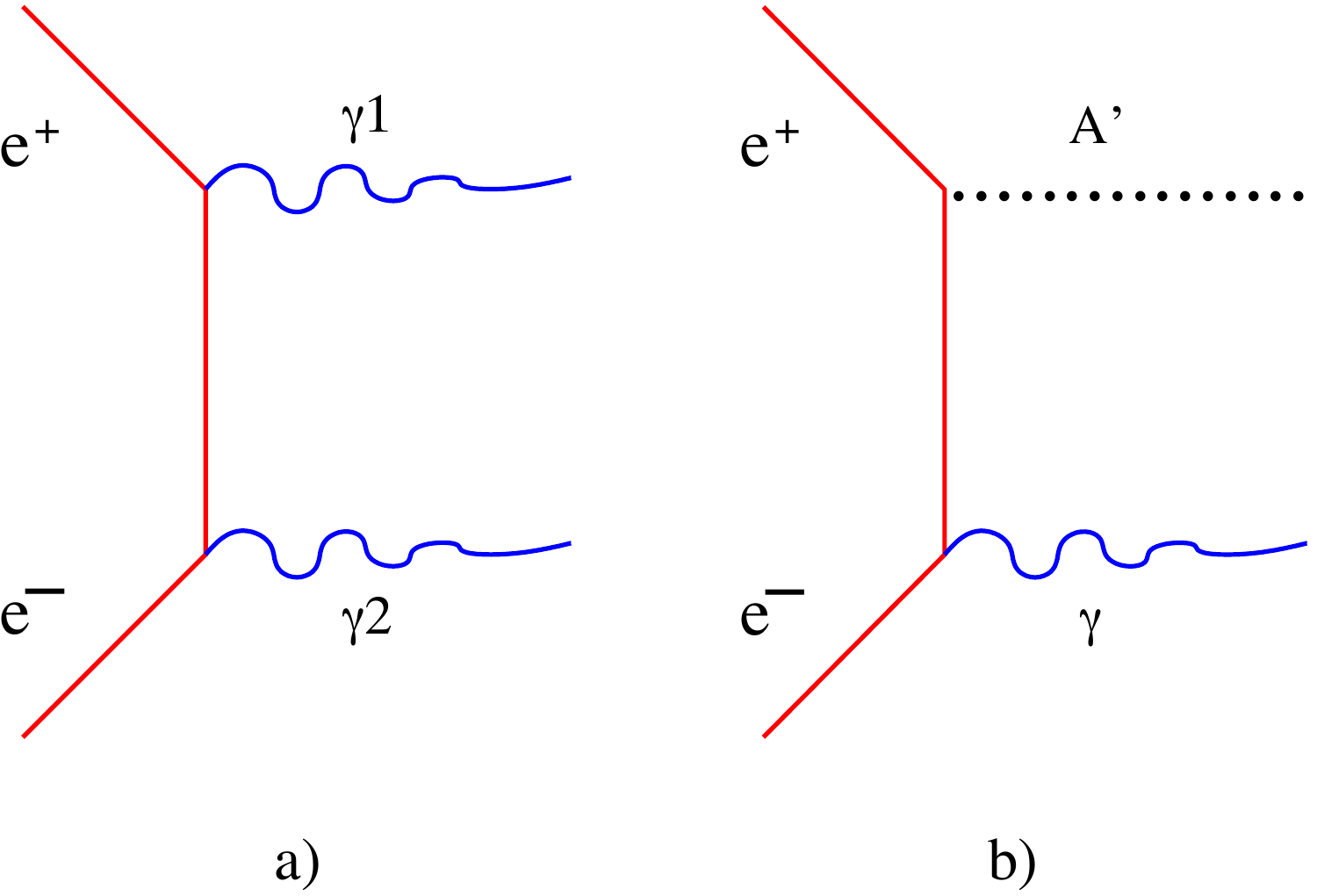}
\hfill
     \includegraphics[width=0.4\linewidth]{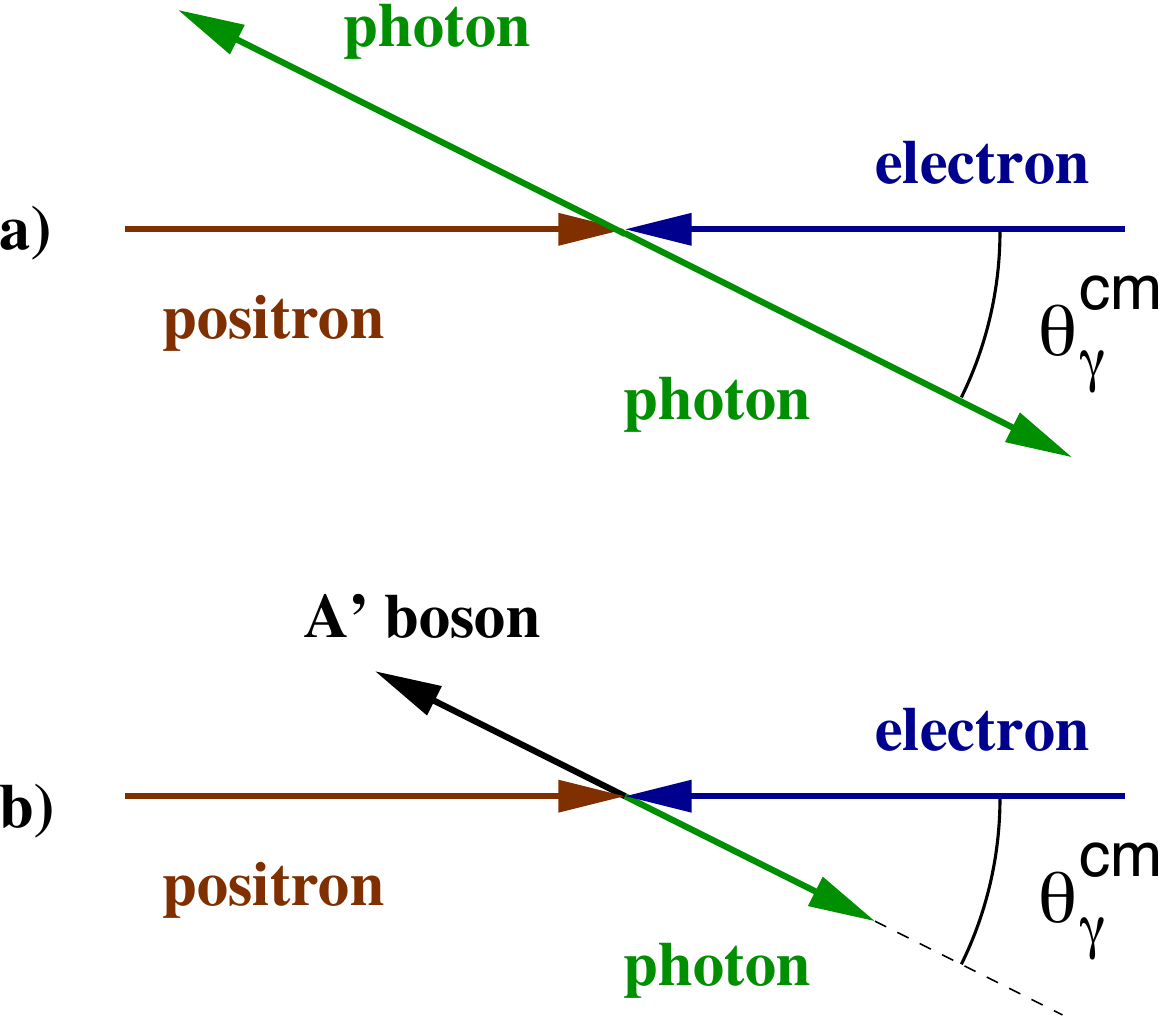}
\caption{
\label{fig:two-body}
The diagrams of a) two-photon annihilation, b) the \AB$-\gamma$ production.
and 
the kinematics of a) $e^+e^- \rightarrow \gamma + \gamma$ and 
b) $e^+e^- \rightarrow {\text A}^\prime + \gamma$ reactions.}
\end{figure}

Two-photon annihilation is a dominant process of high-energy photon 
production in $e^+e^-$ collisions at a cms energy of a few tens of MeV.
Two reactions, depicted in the left panel of Fig.~\ref{fig:two-body}, 
are two-photon annihilation and the production of an exotic \AB. 
The kinematics for the two-body final state is shown in the right panel
of Fig.~\ref{fig:two-body}.
The energy in the center of mass system 
$\sqrt{s} \,=\,  \sqrt{2m^2 \,+\, 2E_+ m} $,
where $m$ is the electron mass and $E_+$ the positron energy,  
and the emission angle of the final photon $\theta_\gamma$ 
with respect to the direction of the positron beam
define the value of the photon energy $E_\gamma$.
In the case of two-photon production: 
$E_{\gamma (\gamma\gamma)}^{lab} \approx E_+ (1 \,-\, \cos\theta_\gamma^{cm})/2$.
In the case of \AB~production: 
$E_{\gamma(A'\gamma)}^{lab} \,=\, E_{\gamma (\gamma\gamma)}^{lab}
\cdot (1 \,-\, M_{A'}^2/s)$.
The kinematic boost from the cm system to the lab leads to a larger photon
energy in the forward direction, which helps the measurement of 
the photon energy.
The large variation of the photon energy with the photon angle 
in the lab system provides an important handle on the systematics. 

\begin{figure}[htb]
\includegraphics[width=0.48\textwidth]{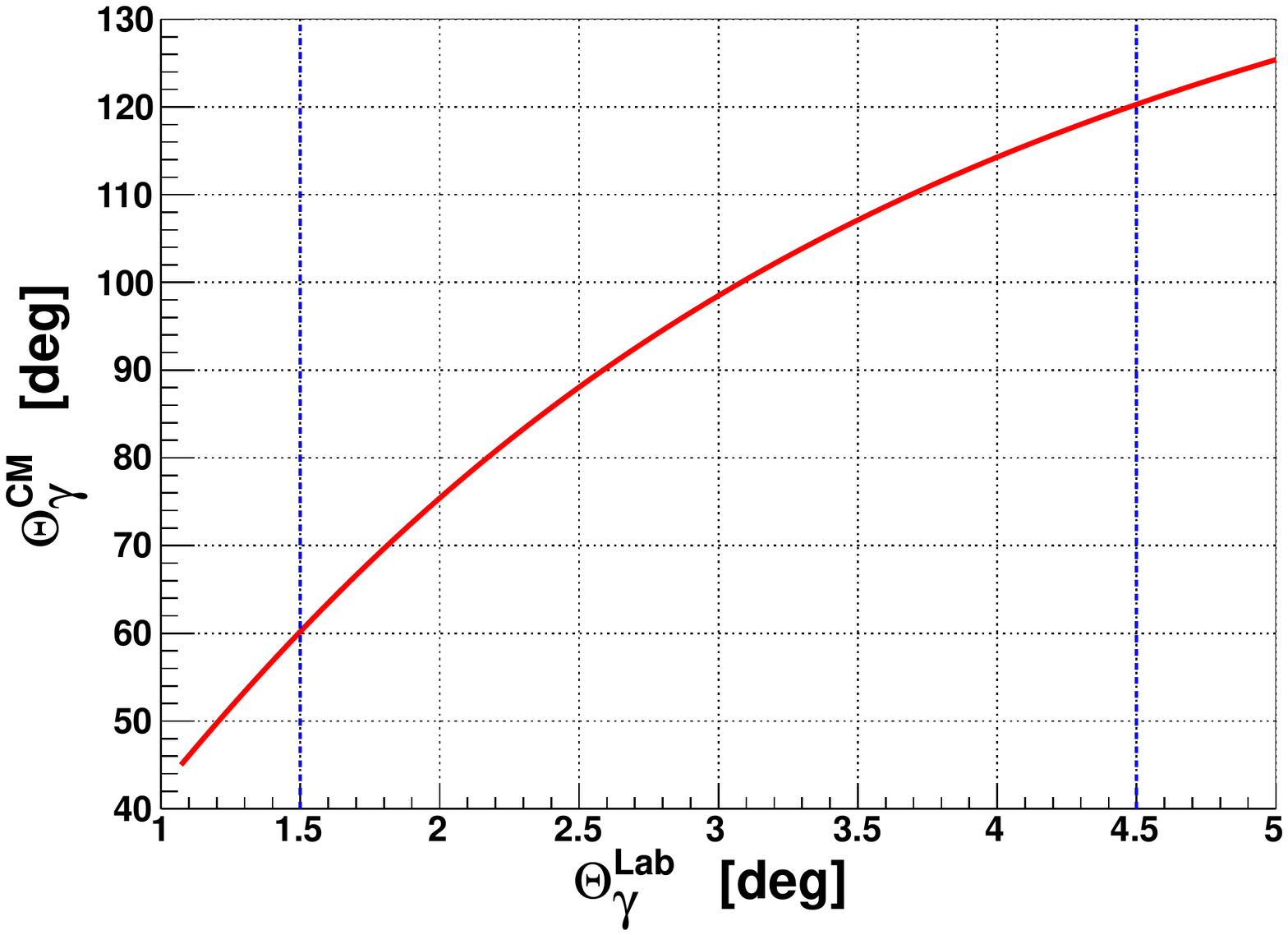}\hfill
\includegraphics[width=0.48\textwidth]{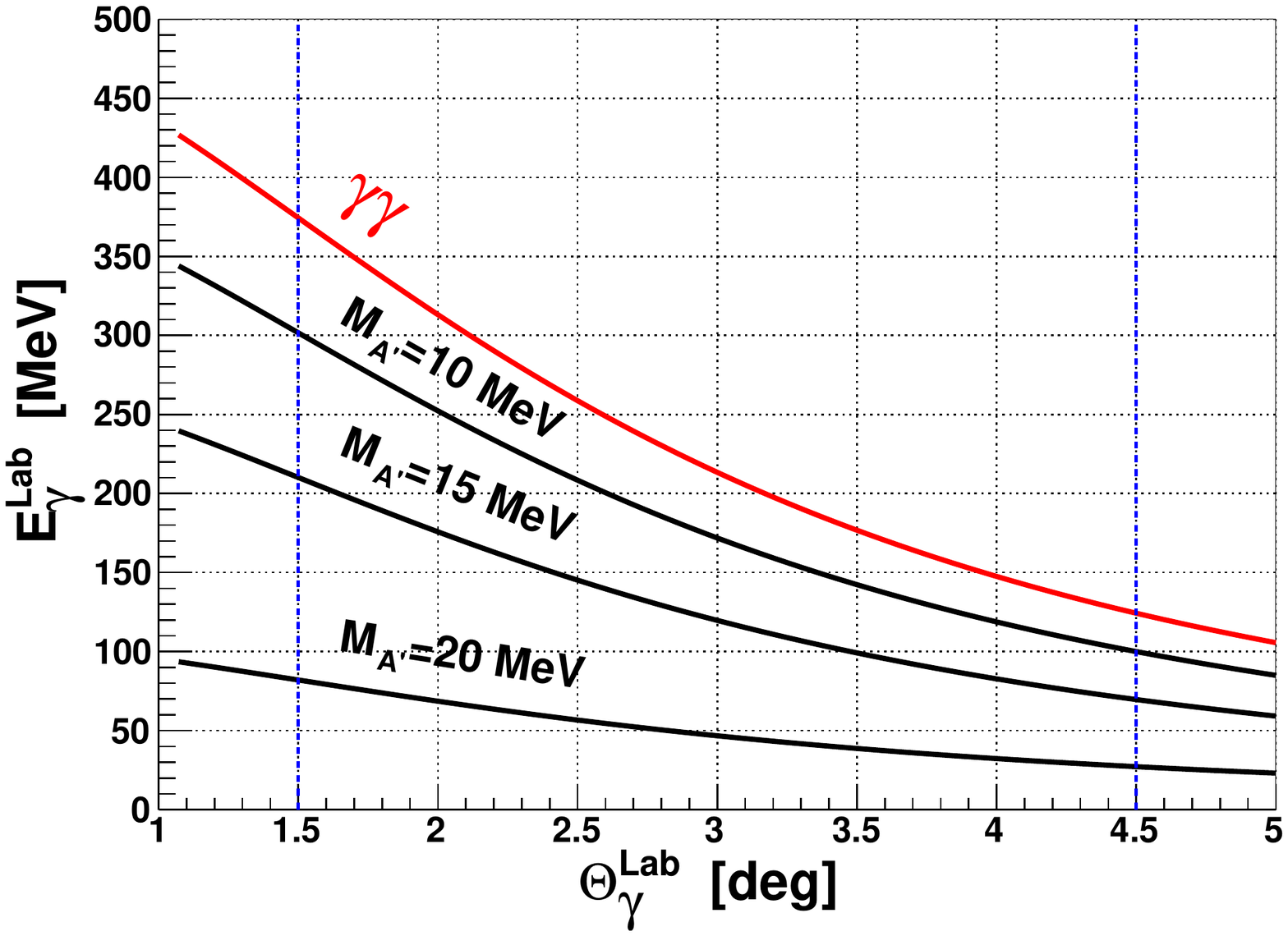}
\caption[]{\label{fig:kinplots}
Kinematic correlations for positron-electron annihilation at E$_+=500$~MeV.
Left panel: photon polar angle in CM frame vs. that in Lab frame for two--photon 
annihilation. Right panel: photon energy vs. its polar angle for 
$e^+e^-\rightarrow \gamma\gamma$ and for $e^+e^-\rightarrow \gamma \text{A}^\prime$.
Dotted vertical lines indicate the range covered in the proposed measurements.
}
\end{figure}

Figure \ref{fig:kinplots} shows some correlations between kinematic variables for
the proposed setup at \V3.

The energy spectrum of the photons from the annihilation process in the lab frame 
is expressed by~\cite{ref:heitler}:
\begin{equation}
\frac{d\sigma}{d\epsilon} = \frac{\pi r^2_e}{\gamma_+-1}
\left\{
  \frac{1}{\epsilon}
\left[
1 -  \epsilon - \frac{2\gamma_+\epsilon -1}{\epsilon(\gamma_++1)^2}
\right]
+ \frac{1}{1-\epsilon}
\left[
\epsilon - \frac{2\gamma_+(1-\epsilon)-1}
{(1-\epsilon)(\gamma_++1)^2}\right]
\right\}
\end{equation} 
where ${\mathbf\epsilon} = E_\gamma^{lab}/(E_++m)$, with 
${\mathbf\epsilon_{min}} =1/2\left[1-\sqrt{(\gamma_+-1)/(\gamma_++1)}\right]$ 
and ${\mathbf\epsilon_{max}} = 1 - \epsilon_{min}$.

The main physical background process producing a single photon, hitting the photon 
detector, is the positron bremsstrahlung. The differential 
cross section of this reaction in the case of a thin hydrogen target is given by the 
following expression \cite{ref:tsai}:
\begin{eqnarray}
\nonumber
\frac{d\sigma_\gamma}{dE_\gamma d\Omega_\gamma}  =  
\frac{4\alpha r^2_e}{\pi}\frac{1}{E_\gamma}
\left\{\frac{2y-2}{\left(1+l\right)^2} 
\right. 
 &+&
\left. 
\frac{12l(1-y)}{\left(1+l\right)^4}
 +\left[\frac{2-2y+y^2}{\left(1+l\right)^2} - \frac{4l(1-y)}{\left(1+l\right)^4}
 \right]
\right. \\
\label{eq:tsai}
& \times &
 \left.
 \left[1+2\ln{\frac{2\gamma_+(1-y)}{y}}-\left(1+\frac{2}{B^2}\right)
 \ln{\left(1+B^2\right)} 
 \right]
\right\}, 
\end{eqnarray}
where $y = E_\gamma/E_+$, $l=\gamma_+^2\theta_\gamma^2$, 
$B= 4\alpha \gamma_+(1-y)/y(1+l)$.

\begin{figure}[htb]
\includegraphics[width=0.48\textwidth]{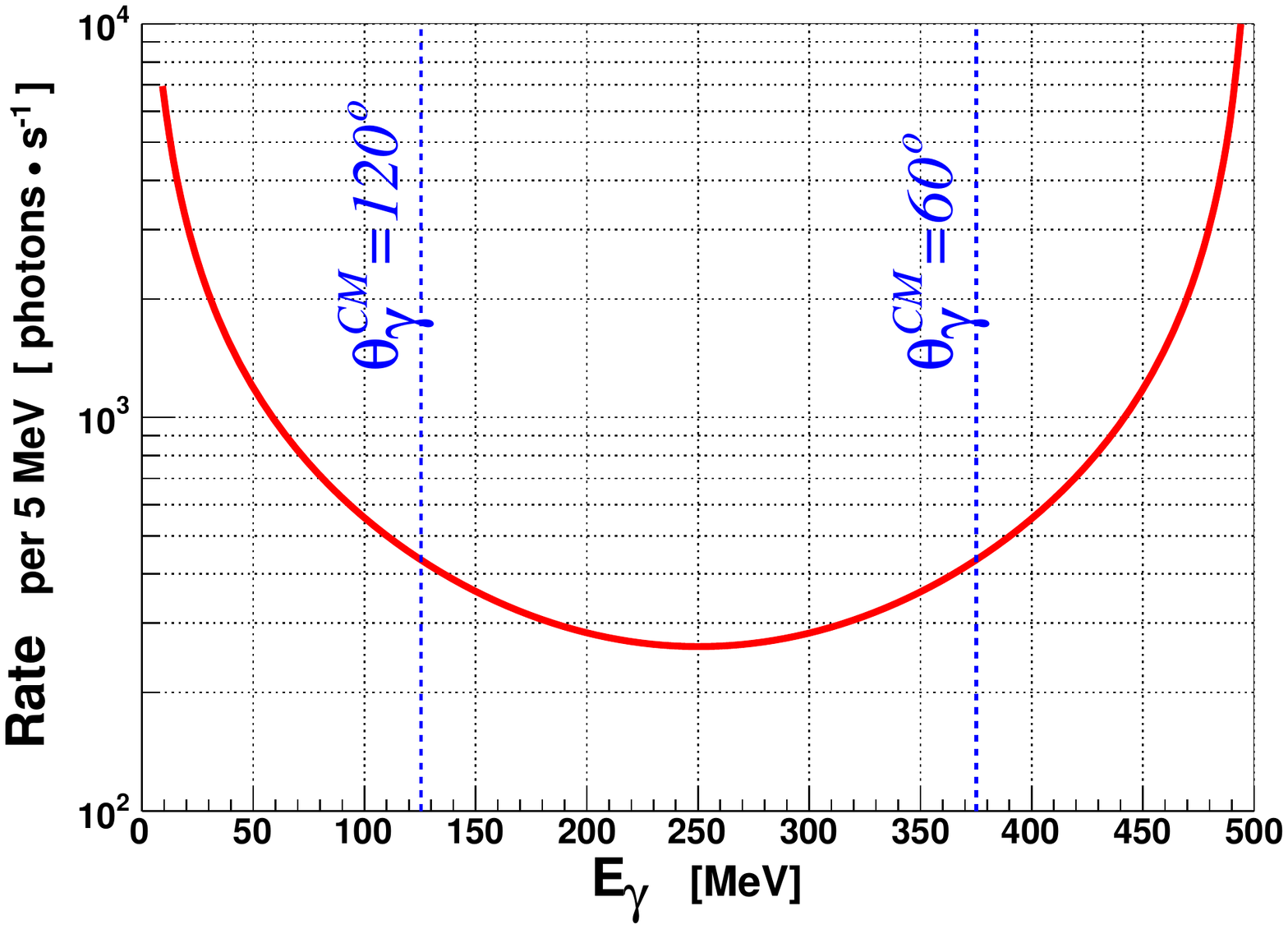}
\hfill
\includegraphics[width=0.48\textwidth]{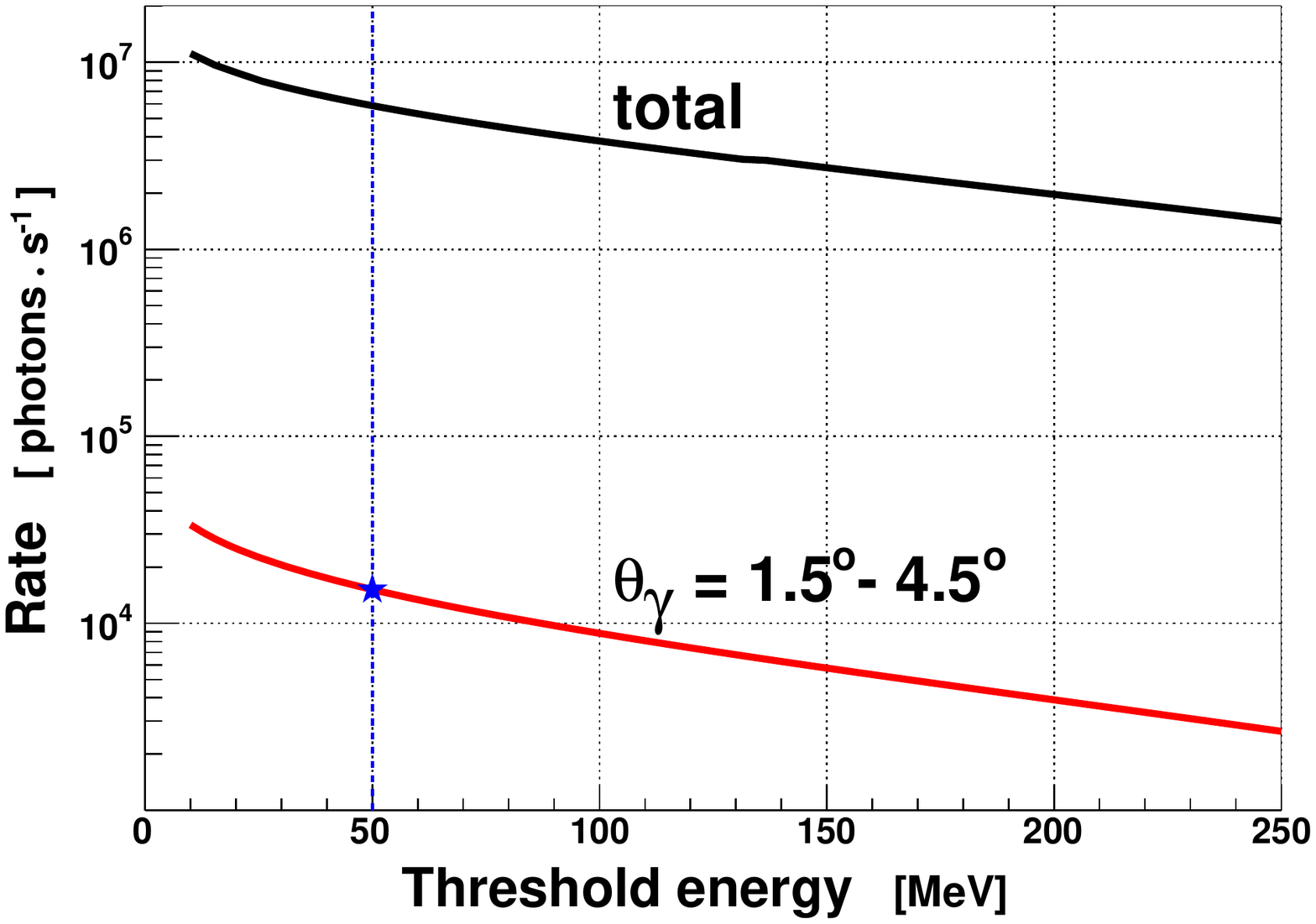}
\caption[]{\label{fig:rates}
Expected background photon rates at beam energy E$_+=500$~MeV and luminosity
$L=10^{32} cm^{-2}s^{-1}$. 
Left panel: from positron-electron annihilation. 
Dotted vertical lines indicate the range covered in the proposed measurements.
Right panel: from positron bremsstrahlung on hydrogen.
At a 50 MeV threshold, the expected rate for the proposed detector configuration is
$1.5\cdot 10^4~s^{-1}$. 
}
\end{figure}

The expected rate of photons from these processes is shown in Fig.~\ref{fig:rates} 

\section{The proposed experimental setup}
\V3 is a buster--ring, operating as an intermediate accelerator/storage ring of electrons 
and positrons for the VEPP-4 collider.     
The internal target is located in one of the two 12-meter-long straight sections of the \V3 ring. 
In the same straight section there are also two RF cavities, four quadrupoles and one sextupole lenses,
and elements of beam injection and extraction. 
The space available for the internal target equipment is 217~cm long. 

The physics program
of the \V3 internal target facility is concentrated on measurements of tensor target asymmetries
in electro- and photo-reactions on a tensor polarized deuteron~\cite{ref:deuteron}.  
Recently, a measurement of the cross-section ratio for the elastic scattering of the positron and 
electron on the proton was carried out \cite{ref:2gamma}.  
The latter demonstrated a reliable 
operation with the positron beam and the internal hydrogen target during a 4-month run at a
luminosity of $\sim 10^{32}$~cm$^{-2}$s$^{-1}$.  
Further improvement of the \V3 performance
is anticipated after the commissioning of the VEPP-5 electron/positron
injection complex is completed~\cite{ref:annrep}.
{ Besides an increase of the positron injection rate,
this new injection scheme will allow up to 18 bunches in \V3,
 which is essential for the reduction of the probability of accidental veto.
}

The layout of the proposed experiment at \V3 is presented in Fig.~\ref{fig:strsec}.
The particle detectors to be used in the proposed experiment and 
the arrangement of components in the internal target area differ
substantially from those used in the previous internal target experiments at \V3. 

\subsection{Beam optics}
In order to allow the photons of the positron-electron annihilation to pass to
the detector without obstruction, a set of dipole magnets will be installed in the beam line.

\begin{turnpage}
\begin{figure*}
  \includegraphics[width=1.0\textheight]{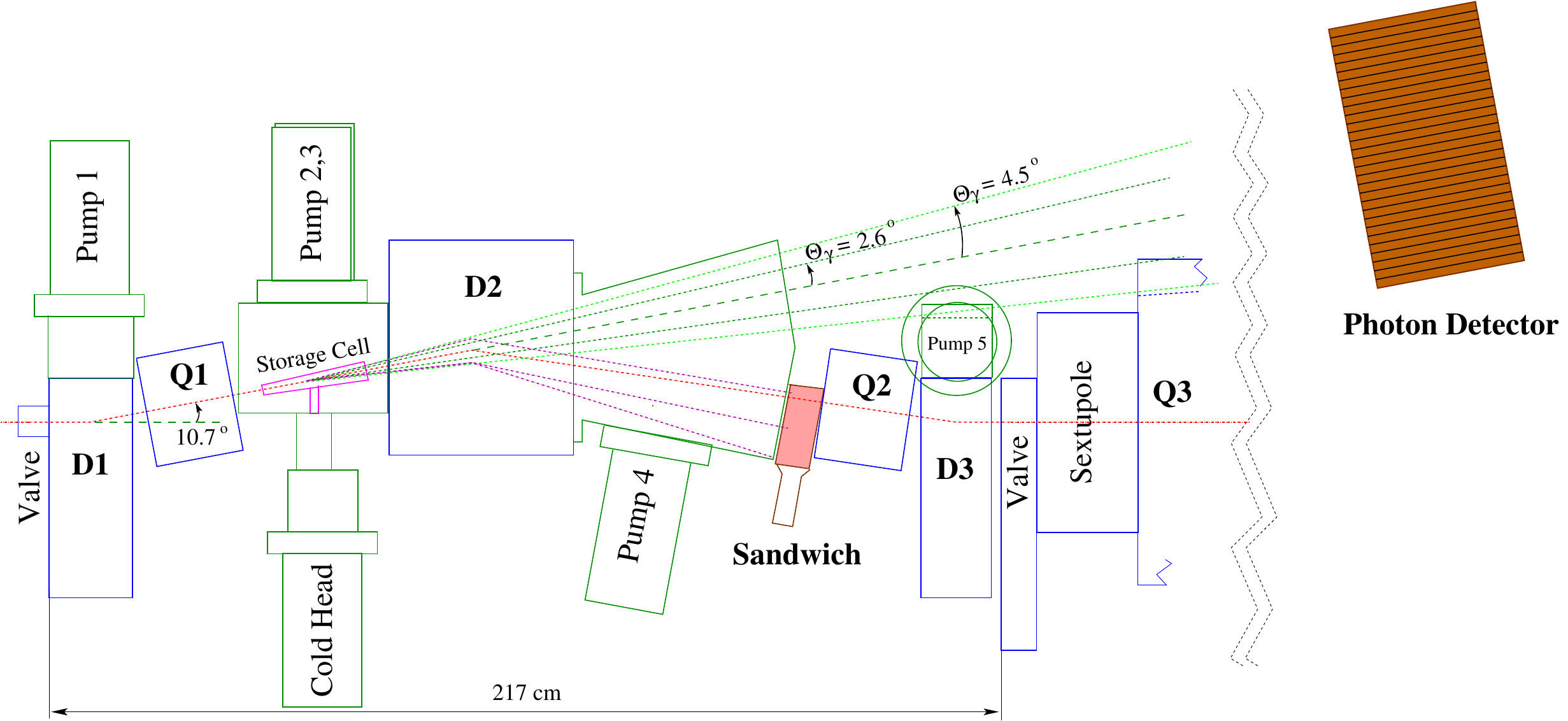}
\caption[]{\label{fig:strsec}
The layout of the proposed experiment at \V3
}
\end{figure*}
\end{turnpage}

Dipoles D1, D2 and D3 (see Fig.~\ref{fig:strsec}) make up a chicane.  The first dipole 
bends the positron beam by an angle of $10^\circ$ toward the ring
center.  The target is placed behind the D1 magnet. The second dipole D2 bends the beam outward from
the ring center by $20^\circ$ and sweeps scattered positrons and other charged particles 
produced in the target away from the photon detector, while photons are flying to the
detector and passing only through a thin beryllium window. 
Finally, the third dipole, D3, is used to rotate the positron beam back to the \V3 beam line.

A similar structure with 3 dipole and 2 quadrupole magnets is now under construction to
be used for the tagging of almost-real photons in
measurements of tensor target asymmetries in photo-deuteron reactions
with a polarized deuterium target \cite{ref:deuteron}.  The D1, D3, Q1 and Q2 magnets
from the Tagger can be directly used in the proposed experiment.  However, the D2 magnet
has to be modified substantially, or even constructed from scratch, because its 
aperture must be significantly larger than is required for the Tagger. 
Furthermore, the sections of vacuum chamber have to be replaced with new ones 
specially designed and manufactured for the proposed experiment.

\subsection{Internal target}
A thin-walled open-ended storage cell cooled to 
$25^\circ\text{K}$ and filled with hydrogen gas will be used as an internal target.
Two additional quadrupoles Q1 and Q2 serve to compress the transverse size of the positron
beam inside the storage cell, thus allowing the use of a
small--opening cell.  Together with
cell cooling this permits us to obtain the required target thickness with a smaller amount of
hydrogen gas injected into the target.  The gas leaks out of the cell ends into 
the ring vacuum chamber and must be pumped out promptly.  Four powerful cryogenic pumps will 
be installed, two in the target chamber, one upstream and one downstream from the target
chamber.

\subsection{Photon detector}
The photon detector can be placed at a distance of between 4~m and 8~m from the target.
The requirements for the detector are:
\begin{itemize}
\item Energy resolution on the level of $\sigma_E/E=5$\% for photons with energy 
$E_\gamma = 100 - 450$~MeV.
\item Angular resolution on a level of $0.1^\circ$.
\item Angular acceptance as defined by a requirement to detect both photons 
from two-photon annihilation:
\begin{itemize}
\item in $\phi$~: either total 2$\pi$, or two symmetrical sectors, e.g. 
$(\phi_1,\phi_2)$ and $(\phi_1+\pi,\phi_2+\pi)$;
\item in $\theta$~: symmetrical range in $\theta_\gamma^{CM}$ around $90^\circ$, e.g.
$\theta_\gamma^{CM}=60^\circ-120^\circ$, which corresponds to 
$\theta_\gamma^{LAB}=1.5^\circ-4.5^\circ$. 
\end{itemize}
\item The detector should be able to sustain a modest photon rate of several hundred
kHz over its whole area. 
\end{itemize} 

The calorimeter for the proposed experiment needs to be constructed.
There are two examples for the calorimeter with suitable parameters: 
\begin{enumerate}
\item The electromagnetic calorimeter of the PRIMEX experiment at JLab \cite{ref:primex}
consists of 1152 lead--tungstate (PbWO$_4$) scintillating crystals of 
$2.05\times 2.05 \times 18$~cm$^3$ size ($20.2X_0$),
surrounded by 576 Lead Glass blocks of $3.8\times3.8\times45$~cm$^3$ size ($10.3X_0$).
In the PRIMEX experiment $\gamma$-quanta with an energy of a few GeV were detected with
resolutions $\delta E/E=2\%$ and $\delta x=1.2$~mm at $E_\gamma=2$~GeV for PbWO crystals
and  $\delta E/E=4.5\%$ and $\delta x=3.7$~mm for Lead Glass blocks.

There is no direct measurement of PRIMEX calorimeter energy and angular accuracy
at photon energy below 1~GeV, but this can be estimated by
extrapolating the energy dependence 
measured for 1--5~GeV. Thus, for $E_\gamma=200$~MeV one obtains :\\
~-- $\delta E/E=5.7\%$ and $\delta x=2.5$~mm for PbWO$_4$ crystals\\
~-- $\delta E/E=13\%$ and $\delta x=11$~mm for Lead Glass blocks.

Although the accuracy of such an estimation is questionable, there is still a clear
indication that the performance of Lead Glass blocks is not adequate for the proposed
experiment, and one should plan to use the inner part of the PRIMEX 
calorimeter, containing the lead--tungstate crystals only. 
That defines the usable area of the photon detector based on the
PRIMEX calorimeter as  $70\times 70$~cm$^2$. Therefore, 
the calorimeter must be installed 
not farther than 4~m from the target 
in order to cover the polar angular range of 4.5$^\circ$. 

As the PbWO$_4$ crystal light yield is highly temperature dependent ($\sim2\%/^o$C 
at room temperature), temperature stabilization of the calorimeter is mandatory.
Throughout the PRIMEX experiment, the calorimeter was operated at
$14\pm0.1^o$C temperature, which was maintained by the circulation of a cool 
liquid around the outer body of the calorimeter assembly.

\item The electromagnetic calorimeter of the CLEO-II detector \cite{ref:cleo}
consists of 8000 CsI(Tl) crystals of $5\times 5\times 30~cm^3$ size ($16.2X_0$).
It is used to measure electron and photon energy in a wide range; therefore, 
a direct measurement of its performance at a photon energy of interest for the proposed
experiment  is available:
$$\delta E/E = 3.8\%  \text{~ and~ } \delta x = 12 \text{~mm} \ \  \text{   for } E_\gamma=180\text{~MeV}$$
One can see that in the energy range of the proposed experiment,  a CsI(Tl)--based 
calorimeter provides better energy resolution but a worse spatial one than 
that based on PbWO$_4$--crystals. Therefore, the CsI(Tl)--calorimeter must be placed
as far as possible from the target, i.e. about 8~m. In this case it would take
about 800 crystals to cover the  required angular range.
A few notes on this detector option should be mentioned:\\
~-- The CLEO-II calorimeter assembly is clearly inappropriate for the proposed experiment,
so a mechanical support must be designed and constructed;\\
~-- CsI(Tl) crystal has a long light emitting time. Therefore, its ability to work at
high background rate is limited. However, due to the relatively small luminosity of the
proposed experiment as well as the high segmentation of the calorimeter, a long 
output pulse seems not to be a problem. Even for crystals covering the lowest
polar angle, the expected rate of background photons is estimated to be at a few 
kHz level for the luminosity of $10^{32}$cm$^{-2}$s$^{-1}$. 
However, it is supposed that the performance of
these crystals will slowly deteriorate as the experiment luminosity is increased.
\end{enumerate}
Both calorimeters could be equipped with front-end electronics and FastBus--based DAQ suitable 
for use in the proposed experiment.

\begin{turnpage}
\begin{figure*}
  \includegraphics[width=1.0\textheight]{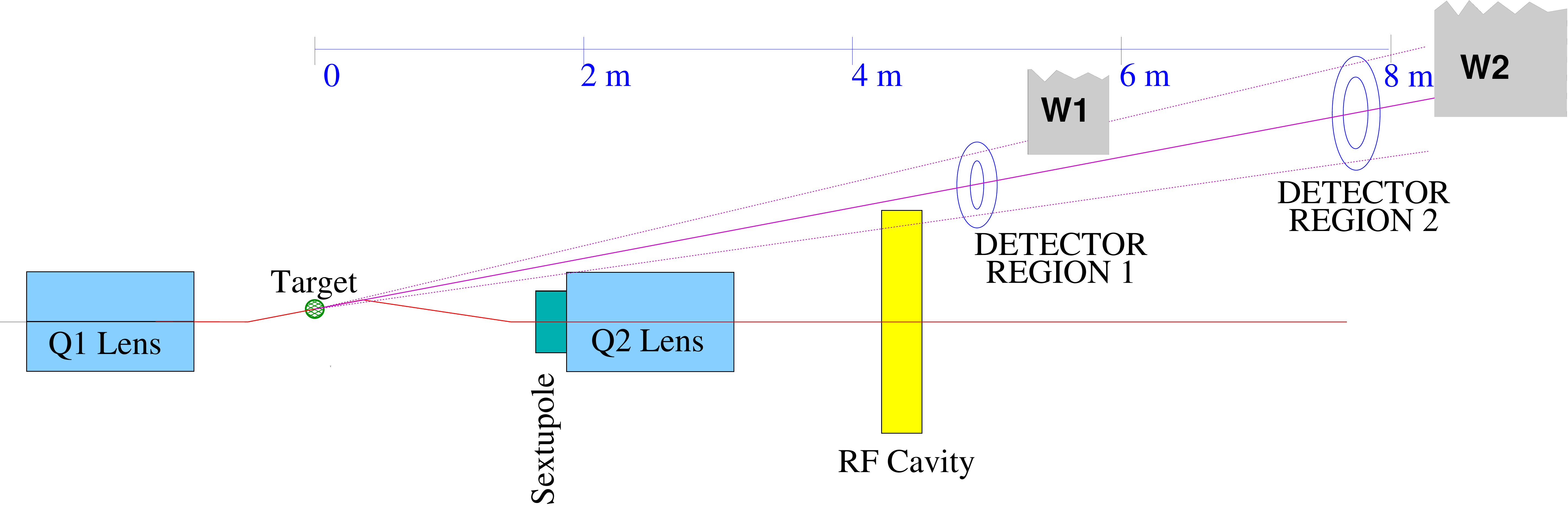}
\caption[]{\label{fig:layout}
Top view at the \V3 area close to the internal target equipment. Two possible locations
of the photon detector are indicated. The RF cavity will be removed; it is not needed for
\V3 when the injection complex is operating. W1 (concrete wall) and Q2 (quadrupole lens)
will be machined around the \V3 median plane to provide a free path for photons 
to the detector.
}
\end{figure*}
\end{turnpage}

In Fig.~\ref{fig:layout} a top view of the \V3 hall in the vicinity of the internal target
equipment and  possible locations of the photon detector are shown.

It seems that the best detector performance would probably be achieved by using a combination of 
PRIMEX and CLEO-II calorimeters, placed at an 8~m distance from the target.  
In such a calorimeter the PbWO$_4$ crystals should be installed at a very 
small polar angle, where photons have the highest energy, thus providing better angular
accuracy, sufficient energy resolution and high rate capability, while the CsI(Tl) crystals 
should cover larger polar angles, where photon rates are much lower, 
providing better resolution at smaller photon energy.

The experiment will require a careful account of the detector responses.
The energy response will be calibrated by using $\gamma-\gamma$
coincidence events produced with the  hydrogen target.  
These data will also provide a detector line shape determination.  
The use of the electron beam instead of the positron beam
provides a way to obtain the ``white'' photon spectra without 
the \AB~signal and the two-photon line. 

\begin{figure}[htb]
\includegraphics[width=0.48\textwidth]{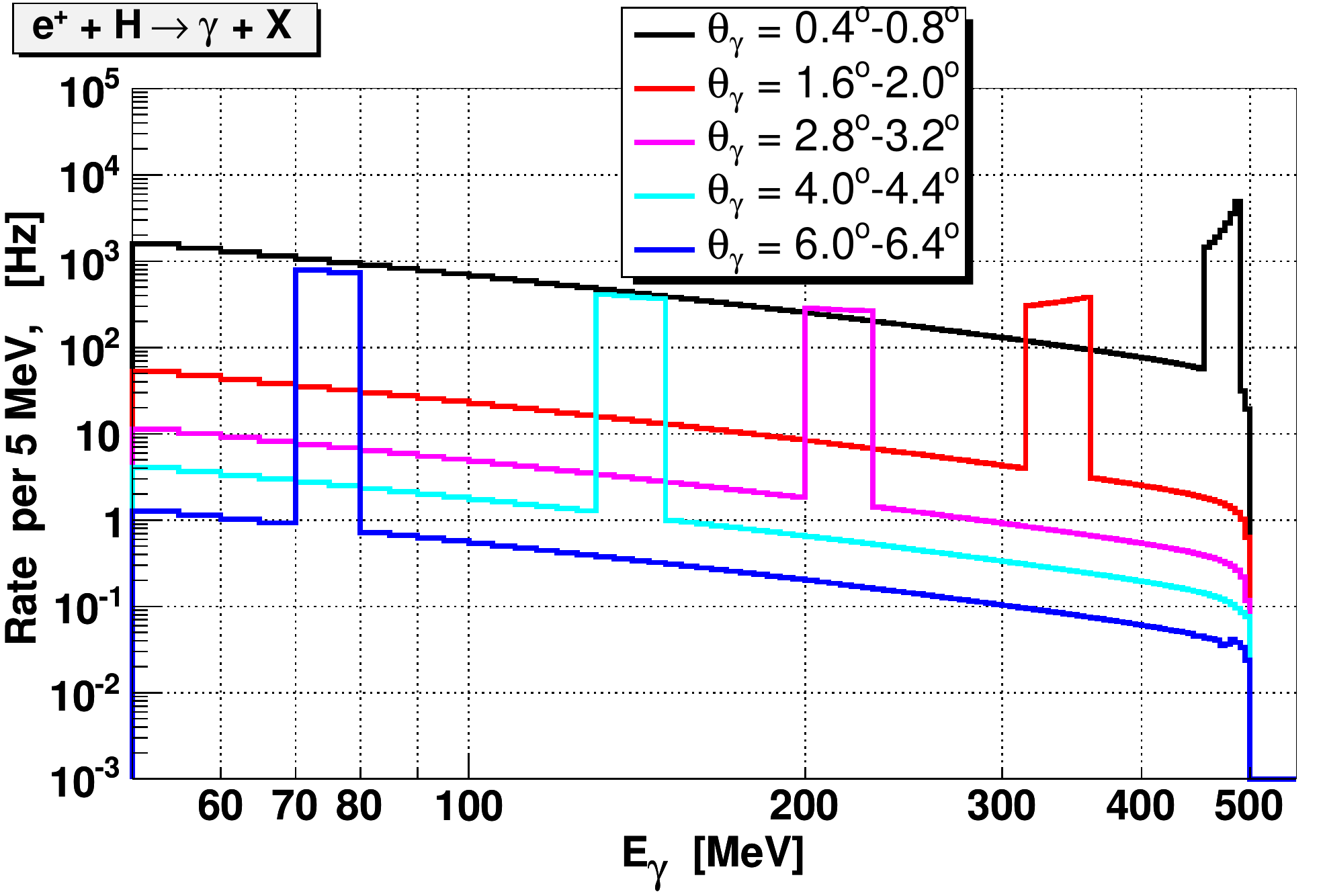}
\hfill
\includegraphics[width=0.48\textwidth]{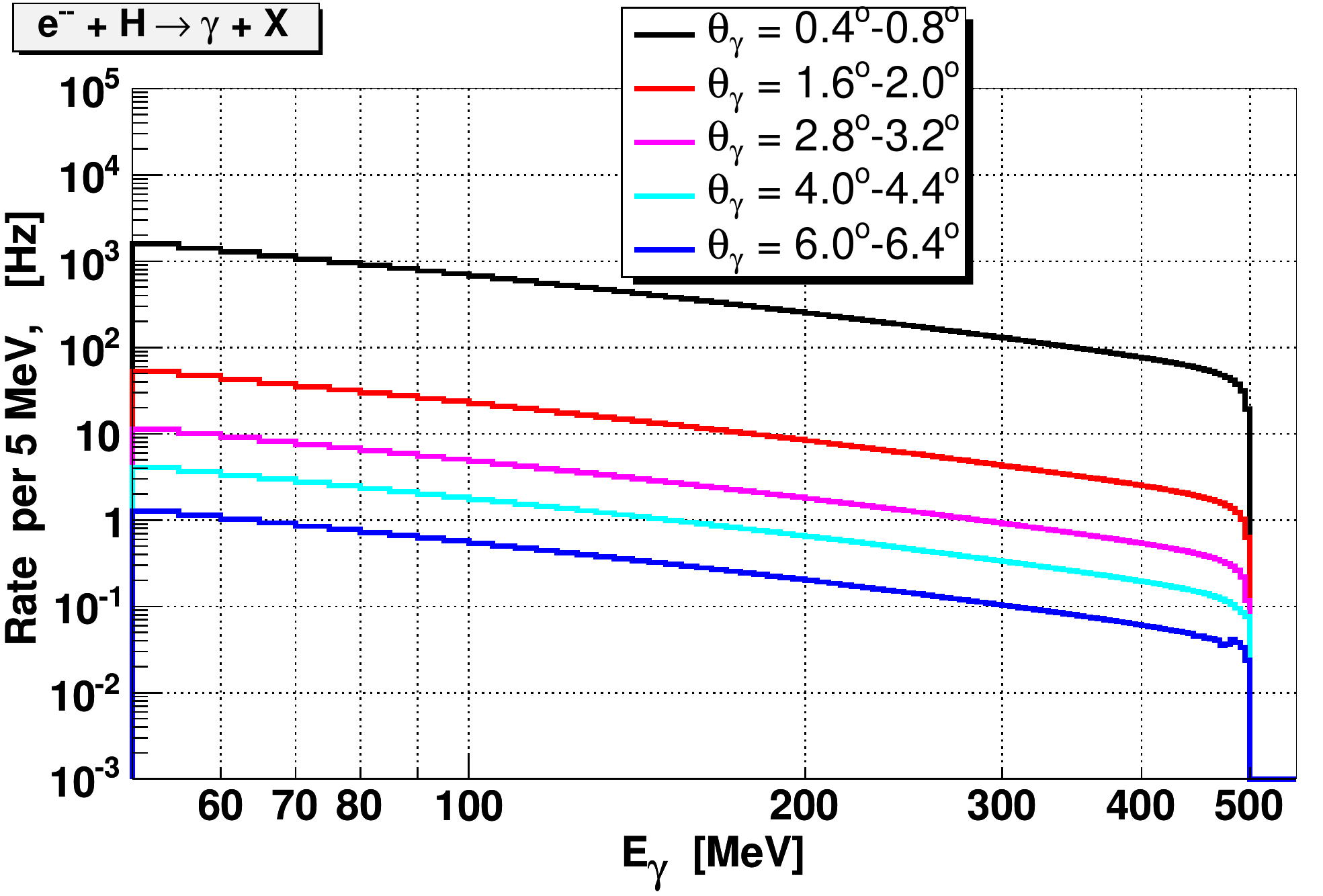}
\caption[]{\label{fig:both}
The photon spectra in the case of a positron beam
(left panel) and an electron beam (right panel) 
incident on an internal hydrogen target with a 
$6.5\cdot 10^{14}$~at/cm$^2$ thickness.
Beam energy is 500~MeV and beam current is 25~mA in both cases.
Bumps on the left panel, whose positions move with the scattering angle,  
are due to the positron-electron annihilation process.
}
\end{figure}

The left panel of Fig.~\ref{fig:both} shows the result of a calculation of 
the photon spectra for the case of an internal hydrogen gas 
target of $6.5\cdot 10^{14}$~at/cm$^2$ thickness ($3\cdot10^{-11}X_0$), 
positron beam current of 25~mA (i.e. a luminosity of $10^{32}$cm$^{-2}$s$^{-1}$) and a 500~MeV 
positron beam energy.
The intensity of the background process in the proposed type of search 
is about 30-100 times below the annihilation process, 
$e^+e^- \rightarrow \gamma \gamma$, whose peak
moves with the scattering angle.
In the case of an electron beam (Fig.~\ref{fig:both}, right panel), 
the energy spectrum is smooth; this will be used for the calibration 
of the detector response.

\subsection{Positron veto counter}
\begin{figure}[htb]
  \includegraphics[width=0.7\linewidth]{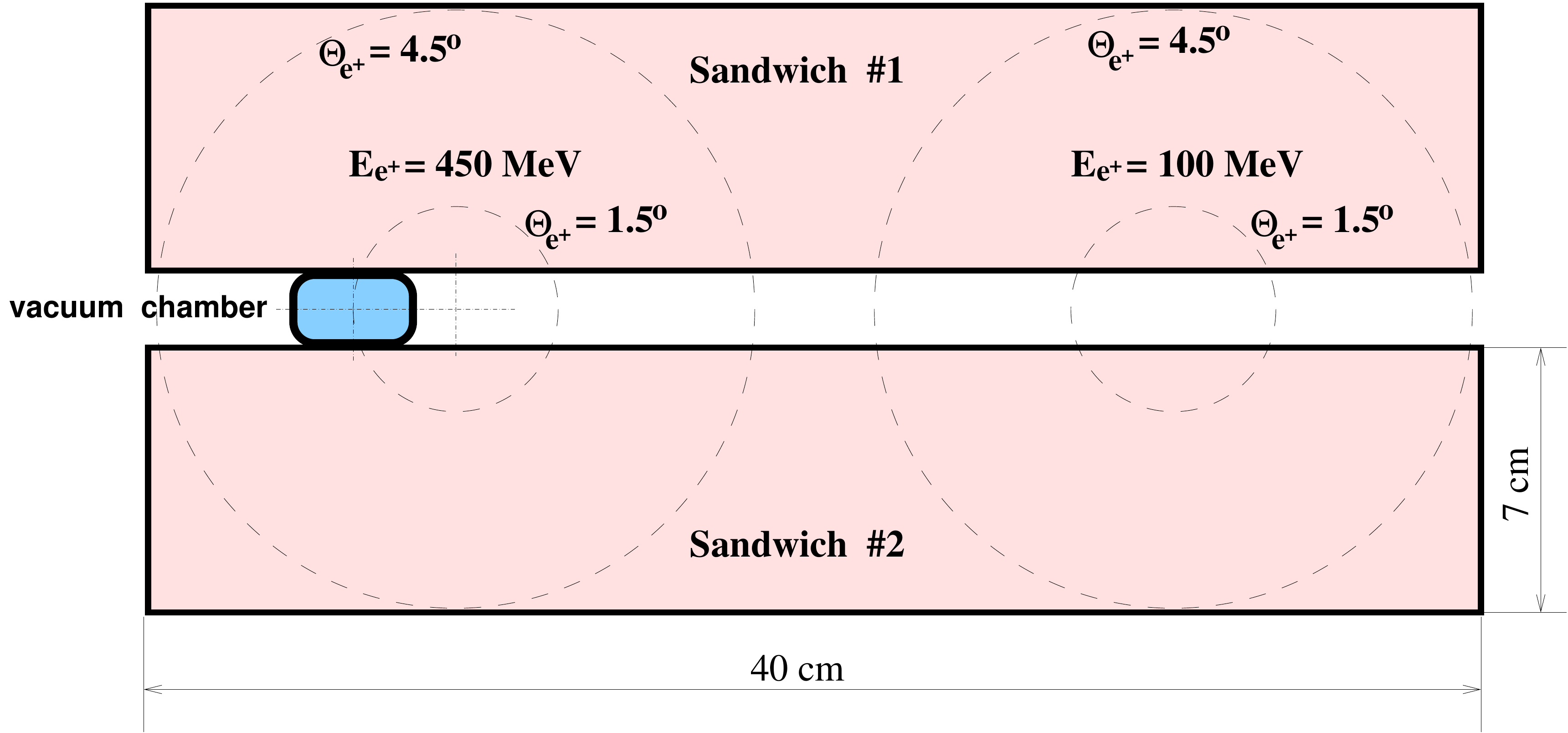}
\caption[]{\label{fig:sandwich}
The layout of the sandwich detectors 
}
\end{figure}

The main single--photon QED background comes from positron bremsstrahlung on hydrogen. 
Since in this process the positron loses energy and is swept out by the D2 dipole
magnet, such background events can be vetoed by detecting the scattered positron.
For this purpose, compact sandwich counters will be installed downstream from the D2 magnet.
To be able to detect as many scattered positrons as possible, the veto detector will
be placed at the largest possible distance from the target--110~cm, right before 
the Q2 quadrupole 
lens, see Fig.~\ref{fig:strsec}. The layout of the sandwich counters is shown
in Fig.~\ref{fig:sandwich}. 

The total rate of positrons emitting a photon with energy above the $E_\gamma=50$~MeV threshold 
can be estimated by integrating Eq.~\ref{eq:tsai}.  For the luminosity
of $L=10^{32}$~cm$^{-2}$s$^{-1}$
the rate is 6 MHz, see Fig.~\ref{fig:rates}.  If \V3 operates in a single-bunch
regime, this would mean that a bremsstrahlung photon with $E_\gamma>50$~MeV is emitted 
every time the beam passes through the target.
The veto counter would be useless at this luminosity if it detects all
positrons including those which emit photons outside the photon detector acceptance.

This is why the veto counter consists of two identical parts placed above and  
below the \V3 vacuum chamber. The vertical gap between the two parts is equal 
to the height of the vacuum chamber $\sim 2$~cm and corresponds to the polar 
angle of 9~mr $ = 9\cdot \gamma_+^{-1}$ for positrons scattered in the vertical plane. 
Due to this gap,  positrons which scatter at a very small angle  will not be detected. 
This dramatically reduces the counting rate of the veto counter and makes 
the anti-coincidence with the photon detector feasible.  
However, a part of the positrons which scatter at the angles of interest 
($\theta = 1.5^\circ-4.5^\circ$)  close to the \V3 median plane will also 
miss the veto counter, and the emitted photons will not be discarded. To avoid this,
a similar horizontal ``band'' in the photon detector will be excluded from 
the trigger and analysis.  The loss in the $\phi$--acceptance due to the gap 
is only about 20\%; it can be compensated for by increasing the experimental luminosity,
which is now not limited by the rate of the veto counter.  

The sandwich detector will be composed of 15 layers of 3~mm thick Tungsten $+$
a 2~mm thick plastic scintillator, for a total thickness of 15$X_0$. 
The expected efficiency
for detecting a positron in an energy range 100--450 MeV is above 98\%. 
Therefore, the events from the bremsstrahlung process will be suppressed by 
a factor of 50.

\subsection{Online trigger and data rate}
The operation of the main on-line trigger will be based on the following logic:
\begin{itemize}
\item threshold on minimum energy deposition in calorimeter, $E_\text{cal}>50$~MeV;
\item veto on positron with energy above 100 MeV detected in the veto counter;
\item veto on maximum energy deposition in the calorimeter, $E_\text{cal}<450$~MeV.
\end{itemize} 
The expected rate of the photon trigger is about 50 kHz. Assuming a conservative value for 
the on-line suppression factor of 10 for combined veto-channels, one obtains a 5 kHz final 
trigger rate, which is reasonably low for the FastBus-based data acquisition system, giving 
a dead time loss below 5\%. With an expected maximum channel occupancy of 20\%  the data
rate will be 4~Mb/s, well below the FastBus specifications. Therefore, the readout
will not contribute to the dead time, provided the event buffering is enabled.

\section{Run time and the projected sensitivity}
The internal hydrogen target and positron beam at \V3 allow a routine operation at 
a luminosity of at least $10^{32}$~cm$^{-2}$s$^{-1}$ with the possibility of increasing by a factor of 5-10.
Considering positron beam energy of 500 MeV, scattering angles 
$\theta^{CM}=90^\circ\pm 30^\circ$ and a luminosity of $10^{32}$~cm$^{-2}$s$^{-1}$, the photon rate
from the annihilation process will be 30~kHz. 
In a {\bf six-month} run the total accumulated statistics will be 
$3.5\cdot10^{11}$ events, assuming 75\% efficiency of time utilization. 

The continuum background is mainly due to positron bremsstrahlung on
hydrogen, which is suppressed by at least a factor of 50 using the veto on detected scattered
positrons and discarding all events with more than one photon in the photon detector.  
Assuming the relative energy resolution of the photon detector 
to be about 5\%, we will use a 15\%-wide energy bin for the search window.  
The background rate in a 15\% photon energy interval after the off-line suppression is
estimated to be 0.08\% of the annihilation photon rate.
Therefore, the statistical uncertainty in the number of background events will be 
$1.7 \cdot 10^4$.  
The \AB~signal with the number of events equal to background 
statistical fluctuation corresponds to
$4.8 \cdot 10^{-8}$ of the annihilation events,
giving a square of coupling constant
$\mathbf{|f_{_{eA'}}|^2 \,=\, 2.2 \cdot 10^{-9}}$.  
Such sensitivity will present up to\ {\bf a 1000-fold}\ improvement
for the limit on the $\mathbf{|f_{_{eA'}}|^2}$  compared 
with the one obtained from the measurement of the electron anomalous
magnetic momentum $g_e$--2, assuming the vector coupling to electron 
and the mass $m_{_{{\text A}^\prime}} \approx 15$~MeV. 

\begin{figure*}[htb]
  \includegraphics[ width=0.8\linewidth]{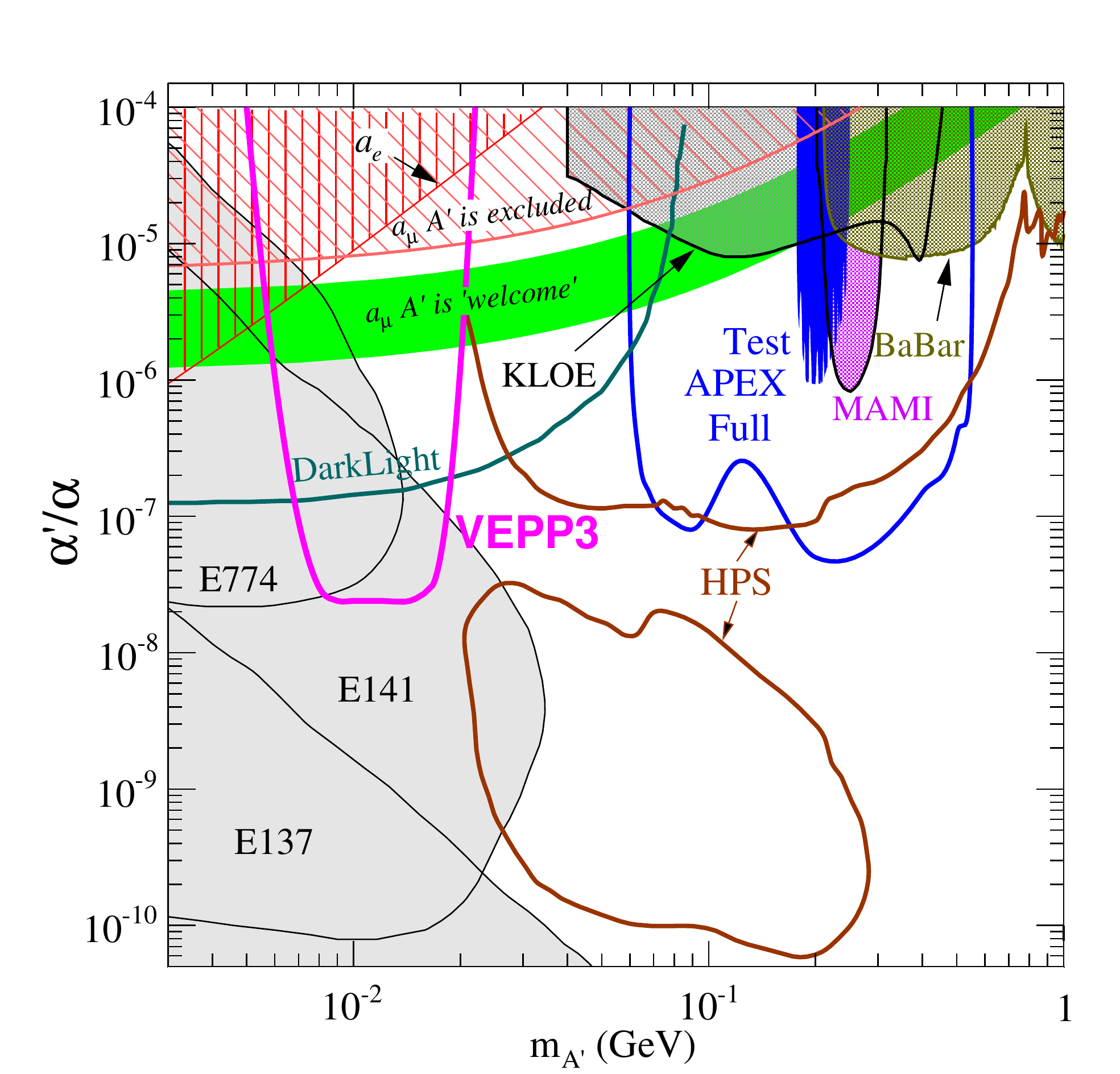}
\caption[]{\label{fig:result}
Existing and projected upper limits of a coupling constant of 
a new boson to lepton vs. its mass.  
The shaded areas are the results of the completed direct searches:
beam dump experiments at SLAC: E137, E141, E774 \cite{bd1, bd2}; 
$e^+e^-$ colliding beam experiments: BaBar \cite{babar}, KLOE \cite{kloe};
fixed--target experiments: MAMI \cite{mainz}, APEX Test run \cite{apres}.
The hatched areas are regions excluded by the results of the measurements of anomalous
magnetic moments of electron and muon \cite{po09}. 
The green band indicates a ``welcome'' area, where the consistency of theoretical
and experimental values would improve to $2\sigma$ or less \cite{po09}. 
Curves show areas of search of other proposed experiments:
Full APEX \cite{apex2}, HPS \cite{HPS}, DarkLight \cite{darklight} and 
the proposed experiment (VEPP3). 
Note that the beam dump experiments 
have  sensitivity only for the processes with the visible decays of the \AB.
Therefore, they don't guarantee a total exclusion of a new boson, 
and the projected \V3 results will provide explicitly new data even in 
the regions already checked by those measurements.  
}
\end{figure*}

Figure~\ref{fig:result} shows a plot of the coupling constant ratio $\alpha'/\alpha$ 
 versus  the mass of the new boson $m_{_{{\text A}^\prime}}$.   
The region  which will be accessible for the search in the proposed experiment is outlined,
together with some other completed and proposed measurements.
(Note that $\alpha'/\alpha = |f_{_{e{\text A}^\prime}}|^2/4\pi\alpha)$.
 
\section{Summary}
We propose a sensitive search for  an exotic \AB~using a missing mass
reconstruction in a positron-electron annihilation, utilizing the \V3 internal target
facility and the VEPP-5 positron/electron injection complex at
the Budker Institute of Nuclear Physics, Novosibirsk, Russia. 

The key features of the proposed measurement are:
\begin{itemize}
\item 
the missing mass method; no assumptions about decay modes of the \AB~are required;
\item the mass range for the proposed search is 5-20 MeV, which is not accessible in
most other proposed fixed-target or colliding-beam approaches;
\item relatively low experimental luminosity ($\sim10^{32}$cm$^{-2}$s$^{-1}$) allows 
the use of available detectors and conventional data acquisition;
\item the use of a veto-detector for scattered positrons and a symmetric 
(in $\phi_\gamma$ and $\theta^{\text{\tiny{CM}}}_\gamma$) photon detector permits a 
suppression of the QED background by a factor of 50-100, resulting in the
increase of the search sensitivity by one order of magnitude.    
\end{itemize} 

\noindent
The projected sensitivity for the square
of the coupling constant of the \AB~to electron is $|f_{e{\text A}^\prime}|^2=1.1\cdot 10^{-8}$
at $m_{_{{\text A}^\prime}}=15$~MeV  
with a signal-to-noise ratio of five to one.

\section{Acknowlegements}

We thank Rouven Essig for productive discussions and helpful feedback.
This work was supported in part by US DOE 
and by the Ministry of Education and Science of the Russian Federation.
Jefferson Science Associates, LLC, operates Jefferson Lab for the US
DOE under US DOE Contract No. DE-AC05-060R23177.

\end{document}